\documentstyle[12pt]{article}

\begin{document}


\def\a{\alpha}
\def\b{\beta}
\def\c{\varepsilon}
\def\d{\delta}
\def\e{\epsilon}
\def\f{\phi}
\def\g{\gamma}
\def\h{\theta}
\def\k{\kappa}
\def\l{\lambda}
\def\m{\mu}
\def\n{\nu}
\def\p{\psi}
\def\q{\partial}
\def\r{\rho}
\def\s{\sigma}
\def\t{\tau}
\def\u{\upsilon}
\def\var{\varphi}
\def\w{\omega}
\def\x{\xi}
\def\y{\eta}
\def\z{\zeta}
\def\D{\Delta}
\def\G{\Gamma}
\def\L{\Lambda}
\def\F{\Phi}
\def\P{\Psi}
\def\S{\Sigma}

\def\o{\over}
\def\beq{\begin{eqnarray}}
\def\eeq{\end{eqnarray}}
\newcommand{\gsim}{ \mathop{}_{\textstyle \sim}^{\textstyle >} }
\newcommand{\lsim}{ \mathop{}_{\textstyle \sim}^{\textstyle <} }

\def\IJMP{Int.~J.~Mod.~Phys. }
\def\MPL{Mod.~Phys.~Lett. }
\def\NP{Nucl.~Phys. }
\def\PL{Phys.~Lett. }
\def\PR{Phys.~Rev. }
\def\PRL{Phys.~Rev.~Lett. }
\def\PTP{Prog.~Theor.~Phys. }
\def\ZP{Z.~Phys. }


\baselineskip 0.7cm

\begin{titlepage}
\begin{flushright}
UT-900
\\
July, 2000
\end{flushright}

\vskip 1.35cm
\begin{center}
{\large \bf
Quintessential Brane and the Cosmological Constant
}
\vskip 1.2cm
Ken-Iti Izawa
\vskip 0.4cm

{\it Department of Physics and RESCEU, University of Tokyo,\\
     Tokyo 113-0033, Japan}

\vskip 1.5cm

\abstract{
We consider a quintessential sector realized
on a hidden brane with a tiny warp factor,
which implies a small value of the effective four-dimensional
cosmological constant through an equation of motion
for the quintessential scalar.
}
\end{center}
\end{titlepage}

\setcounter{page}{2}


The minuscule scale characterizing
the physics of the cosmological constant
\cite{Wei,Car}
or quintessence
\cite{Car,Bin}
\footnote{For an investigation of quintessence in brane cosmology,
see Ref.\cite{Gon}.}
provides the largest hierarchy among fundamental physical parameters.
This hierarchy suggests that the sector responsible
for such a minuscule scale may be separate from our visible
sector, in which the standard model of elementary particles resides.

In this paper, we consider the possibility that this separate sector
is realized on a quintessential brane which is
hidden from the standard sector on another brane or in a bulk.
Namely, we consider a quintessential sector realized
on a hidden brane with a tiny warp factor,
which implies a small value of the effective four-dimensional
cosmological constant through an equation of motion
for the quintessential scalar.

Let us consider a scalar field $\f$ on a hidden 3-brane coupled to
gravity induced from the bulk spacetime
whose action is given by
\beq
 S = \int \! d^4x \sqrt{-g}\left({1 \o 2}g^{\m \n}\q_\m \f \q_\n \f
   - V(\f) + {1 \o 2}U(\f)R\right),
 \label{SL}
\eeq
where $g_{\mu \nu}$ denotes the four-dimensional metric induced
on the brane from the bulk gravity.%
\footnote{The potential $V(\f)$ may be flat due to (super)symmetry
at a large scale in terms of the fundamental metric $g_{\mu \nu}$.}

Under a warped compactification of the higher-dimensional bulk theory,
the four-dimensional metric ${\bar g}_{\mu \nu}$
visible in our world is related to the induced metric
through a warp factor:
$g_{\mu \nu}=a^2{\bar g}_{\mu \nu}$
\cite{Rub,Ran,Hor}.
Then the sector given by Eq.(\ref{SL}) amounts to
\beq
 S = \int \! d^4x \sqrt{-{\bar g}}
   \left({1 \o 2}{\bar g}^{\m \n}\q_\m {\bar \f} \q_\n {\bar \f}
   - {\bar V}({\bar \f}) + {1 \o 2}{\bar U}({\bar \f}){\bar R}\right)
\eeq
in the effective four-dimensional theory,
where ${\bar \f}=a\f$, ${\bar V}({\bar \f})=a^4V(\f)$
and ${\bar U}({\bar \f})=a^2U(\f)$.
Note that ${\bar V}({\bar \f})$ and ${\bar U}({\bar \f})$
do not directly yield the vacuum energy and the gravitational
constant in our universe. There are contributions (of larger size)
from the standard and other sectors
on other branes or in a bulk, which are not explicit here.

When the warp factor $a^2$ is very small,%
\footnote{The warp factor can be exponentially small
\cite{Ran}
with an AdS slice in an $S^1/Z_2$ orbifold
\cite{Ran,Hor},
for instance. Such a concrete example with its interpretation
as spacetime inflation will be given elsewhere.}
the potential ${\bar V}$ can be quintessentially small
without fine tuning,
provided that the potential $V$ in terms of the fundamental
metric $g_{\mu \nu}$ is smaller than the fundamental scale.
In fact, the equation of motion for the constant scalar field is given by
\beq
 {1 \o 2}{\bar U}'{\bar R}={\bar V}',
\eeq
that is,
\beq
 {1 \o 2}U'{\bar R}=a^2 V',
\eeq
which may allow for realization of the quintessential
vanishing of the four-dimensional cosmological constant
\cite{Iza}
with the values of $V'/U'$ that are not too large.
We note that this is possible since the four-dimensional metric
induced on the brane from the bulk spacetime has meaning that is
independent of the field-dependent Weyl rescalings on the brane
in the warped compactification.


\end{document}